\documentclass{ECOC-author}

\begin{document}

\title{A MODEL FOR THE GENERALIZED DROOP FORMULA}

\author{A. Bononi\ad{1*}, J.-C. Antona\ad{2}, A. Carbo M\'eseguer\ad{2}, P. Serena\ad{1}}

\address{\add{1}{Department of Engineering and Architecture, Universit\`a di Parma, Italy}
\add{2}{Alcatel Submarine Networks, Villarceaux, France}
\email{* alberto.bononi@unipr.it}}


\begin{abstract}
We provide a new analytical model that fully justifies the recently
disclosed Generalized Droop Formula of the
nonlinear signal-to-noise (SNR) ratio in very-long submarine links
with power-mode amplifiers, and show its relation with the Gaussian-Noise
model SNR.
\end{abstract}

\maketitle

\section{Introduction}

It is well known that amplified spontaneous emission (ASE) induces
a droop of the desired signal power at the end of a very long submarine
link with end-span power-mode amplifiers, thus reducing the received
signal to noise ratio (SNR) \cite{bergano,subcom,ASN_droop_ofc19}.
The generalized droop formula (GDF) for the received SNR has recently
been heuristically introduced to also account for the effect of nonlinear
interference (NLI) on droop \cite{ASN_droop_ofc19,asn_suboptic}. It is the purpose
of this paper to provide a solid mathematical model that fully justifies
the GDF and allows a direct comparison with the SNR provided by the
Gaussian Noise (GN) model \cite{poggioPTL11,grellierOptExp}.

\section{Droop in power-mode amplifiers}

We begin with the droop induced by ASE only. Consider the transmission of a single optical channel (normally a wavelength division multiplexed (WDM) comb: amplifier gain is assumed to be flat over the comb) of power $P$ along a chain of $N$ \emph{identical}
fiber spans. The $k$-th span has fiber span loss (in linear units)
$\mathcal{L}<1$. The span is followed by an optical amplifier operated
in power mode that makes the line transparent, i.e., its output power
is again $P$. Since $P\mathcal{L}$ is the power entering the amplifier,
the gain $G$ is such that $P=(P\mathcal{L}+\mu_{a})G$, where $\mu_{a}=h\nu\,F_{N}\,B_{amp}$
is the equivalent input power of the ASE noise generated within the
amplifier bandwidth $B_{amp}$ ($h\nu$ is the photon energy at the
center frequency $\nu$; $F_{N}$ is the amplifier noise figure, which
for large-enough gain $G$ is independent of the gain itself). Hence
$G=\frac{P}{P\mathcal{L}+\mu_{a}}=\mathcal{L}^{-1}\chi_{a}$ where
\begin{equation}
\chi_{a}\triangleq(1+\frac{\beta}{P})^{-1}\label{eq:chiA}
\end{equation}
is the gain droop due to ASE, with $\beta=\mu_{a}\mathcal{L}^{-1}$
the per-span output ASE that would be generated in absence of droop.
~

What is happening is re-interpreted graphically in Fig.
\ref{fig:charts}(top). The droop term $\chi_{a}<1$ pops out to attenuate
the signal power $P$ entering the $k$-th line+amplifier span (sum
of desired signal $P_{s}(k-1)$ and cumulated ASE $P_{a}(k-1)$) down
to $\chi_{a}P$ in order to make room for the locally-generated ASE
$\delta P_{a}=\mu_{a}G\equiv\beta\chi_{a}$ and meet the output power
constraint $P$, so that $P\chi_{a}+\delta P_{a}=P$. This is an equation
in the variable $\chi_{a}$ whose solution is (\ref{eq:chiA}).

Note that the expressions of gain and droop remain the same if we
interpret $P$ as the per-channel power in a $C$-channel single-mode fiber WDM system
and $\mu_{a}$ is calculated on the per-channel bandwidth $B_{amp}/C$\footnote{\textcolor{red}{It is tacitly assumed that WDM signal and ASE occupy the same bandwidth.}}. Similarly, if propagation is on multi-mode fibers  where modal loss and amplification are identical for all modes, then again  $P$ is interpreted as the per-mode per-channel power.
Thus for every channel and mode, the $\textrm{SNR}$ at the output of the $N$-span chain
is obtained by noting that $P\chi_{a}^{N}$ is the received signal
power, and thus $P(1-\chi_{a}^{N})$ must be the ``droop aware''
received ASE power, so that one gets \cite{ASN_droop_ofc19,asn_suboptic}: $\textrm{SNR}=1/(\chi_{a}^{-N}-1)$.

\subsection{Droop induced by NLI}

NLI-induced signal depletion, or droop, was already empirically tackled
in \cite{poggio_depletion-2} with gain-mode amplifiers. We show next
that with power-mode amplifiers the treatment of NLI-induced gain
droop can be nicely integrated with the ASE-induced gain droop, if
we assume that the NLI contributions of the various spans are uncorrelated,
as in the incoherent GN model. 

\begin{figure}
 
\centering\includegraphics[width=0.5\columnwidth]{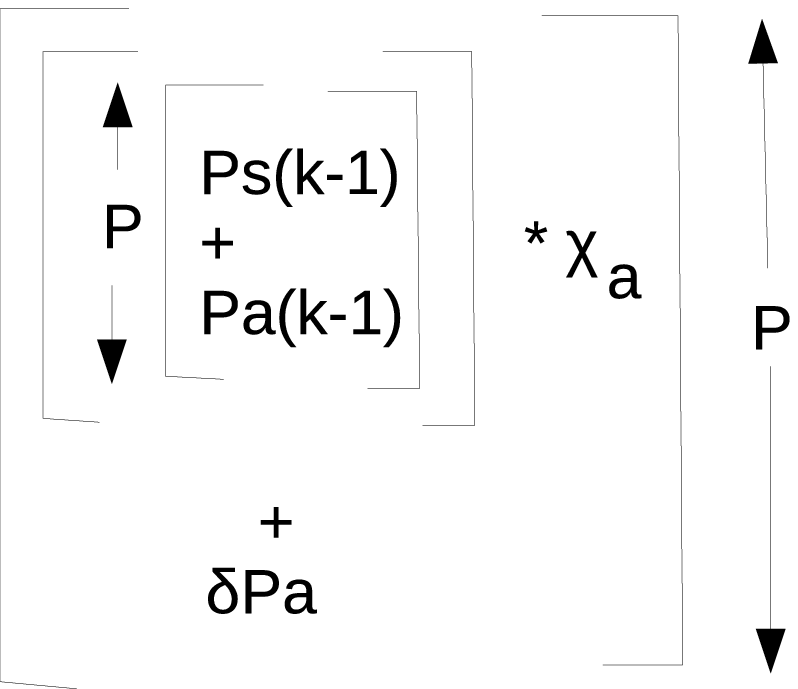}

\includegraphics[width=0.7\columnwidth]{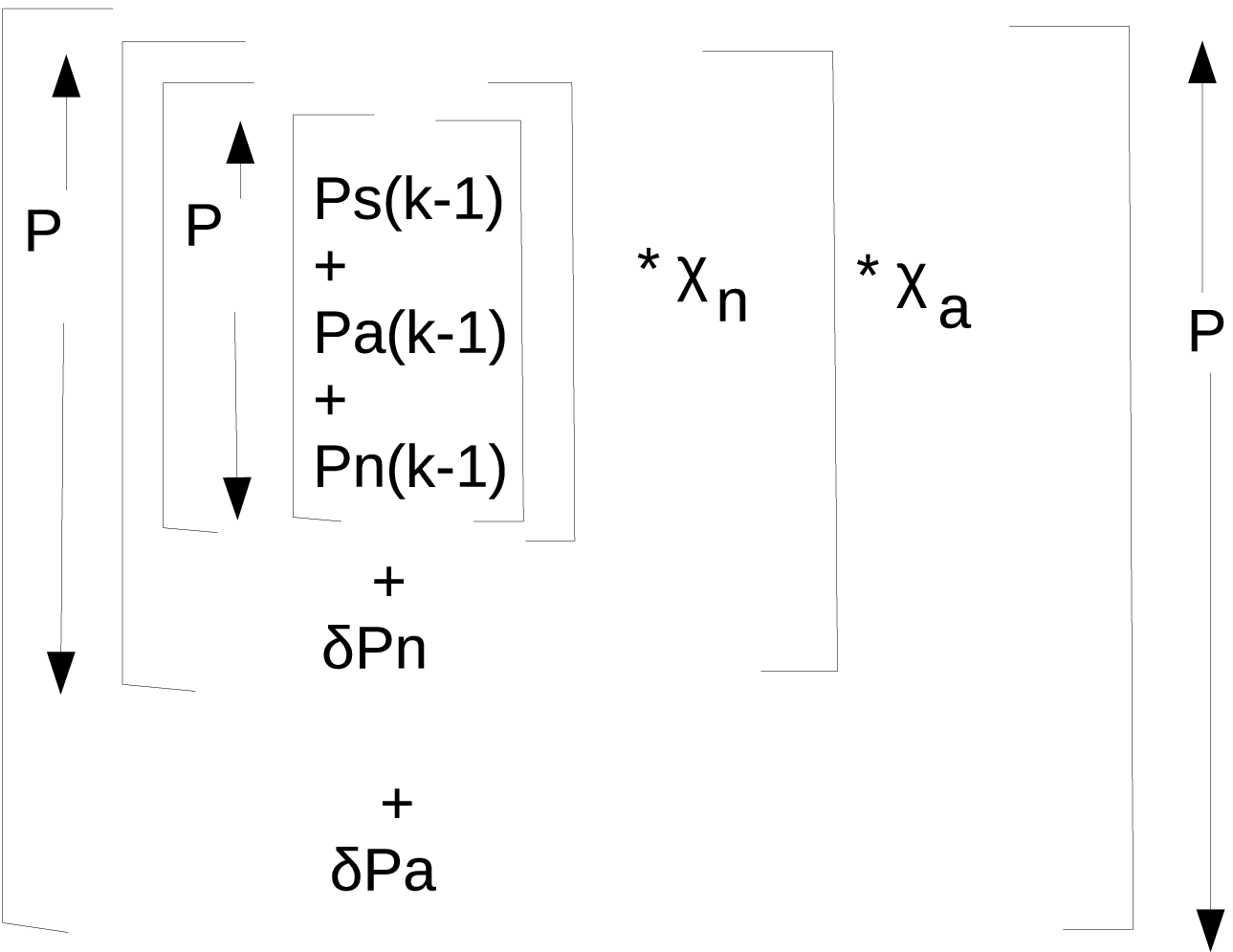}

\caption{\label{fig:charts} Origin of droop:\textbf{ (top)} ASE only. \textbf{(bottom)}
NLI and ASE. (s,a,n)$\rightarrow$(signal, ASE, NLI).}
 
\end{figure}

The logical reasoning goes as depicted graphically in
Fig. \ref{fig:charts}(bottom). The initial powers at output of amplifier
$k-1$ (input of $k$-th fiber span) are $P_{s}(k-1)+P_{a}(k-1)+P_{n}(k-1)=P$.
Nonlinearity operates a redistribution of the power $P$: it generates
NLI as $\delta P_{n}=\alpha_{\textrm{NL}}P^{3}$ \cite{poggioPTL11,grellierOptExp},
where $\alpha_{\textrm{NL}}$ is the per-span NLI coefficient, and rescales
all power components entering the fiber $P_{s}(k-1),P_{a}(k-1),P_{n}(k-1)$
by the same multiplicative scaling factor $\chi_{n}<1$ such that
power is redistributed but otherwise conserved: $P\chi_{n}+\delta P_{n}=P$,
i.e.,
\begin{equation}
\chi_{n}=1-\frac{\delta P_{n}}{P}=1-\alpha_{\textrm{NL}}P^{2}.\label{eq:chiN}
\end{equation}
 Note that \emph{NLI here includes nonlinear interaction of signal
with itself, ASE with itself,
NLI with itself, signal with ASE, signal with NLI, and ASE with NLI}.

Now the story goes as in the ASE-only previous case: the compound
signal (desired plus ASE plus NLI) entering span $k$ sees a gain
of $G\triangleq\mathcal{L}^{-1}\chi_{a}$ (which defines the ASE-induced
droop $\chi_{a}\leq1$), hence at output of $k$th amplifier has power
$\chi_{a}P$, which is the ``attenuated output signal component''
at amplifier $k$, to which the amplifier-generated output ASE $\delta P_{a}$
is added in order to form the amplifier total output power $P$:~$\chi_{a}P+\delta P_{a}=P$,
leading to the usual expression of the ASE droop (\ref{eq:chiA}).

The $\textrm{SNR}$ at the output of the $N$-span chain is obtained by noting
that $P\chi^{N}$ is the received signal power, where $\chi=\chi_{a}\chi_{n}$
is the total signal power droop, and $P(1-\chi^{N})$ is the ``droop
aware'' received ASE+NLI power, so that one gets the generalized
droop formula: 
\begin{equation}
\textrm{SNR}_{\textrm{GDF}}=\frac{1}{\chi^{-N}-1}=\frac{1}{\left[\left(\frac{1}{1-\alpha_{\textrm{NL}}P^{2}}\right)\left(1+\frac{\beta}{P}\right)\right]^{N}-1}\label{eq:GDF}
\end{equation}
which coincides with that reported in \cite{ASN_droop_ofc19,asn_suboptic} for
small NLI to signal power ratio $a_{\textrm{NL}}P^{2}$, i.e., within the first-order
perturbative limits of validity of the GN model.

We note that the NLI droop derivation extends verbatim to any power-redistributing
effect in the fiber, such as for instance the guided-acoustic wave
Brillouin scattering (GAWBS) \cite{GAWBS,asn_suboptic}. Considering both NLI and
GAWBS, we find that the \emph{redistribution} droop factor (\ref{eq:chiN})
to be used in the GDF-SNR (\ref{eq:GDF}) becomes $\chi_{n}=1-\alpha_{\textrm{NL}}P^{2}-\gamma_{GAWBS}\ell$,
where $\gamma_{GAWBS}$ (km$^{-1}$) is the GAWBS coefficient and
$\ell$(km) the span length.

The GDF-SNR (\ref{eq:GDF}) should be checked against the GN-model
SNR. The GN model is traditionally derived using gain-mode amplifiers,
whose gain is $G=\mathcal{L}^{-1}$ and neglects droop. The SNR, assuming
incoherent accumulation of NLI, is calculated as
\begin{equation}
\textrm{SNR}_{\textrm{GN}}=\frac{P}{\mu_{A}GN+\alpha_{\textrm{NL}}P^{3}N}=\frac{1}{N}\left(\frac{1}{\frac{\beta}{P}+\alpha_{\textrm{NL}}P^{2}}\right)\label{eq:GN}
\end{equation}

\begin{figure}
\centering\includegraphics[width=0.85\columnwidth]{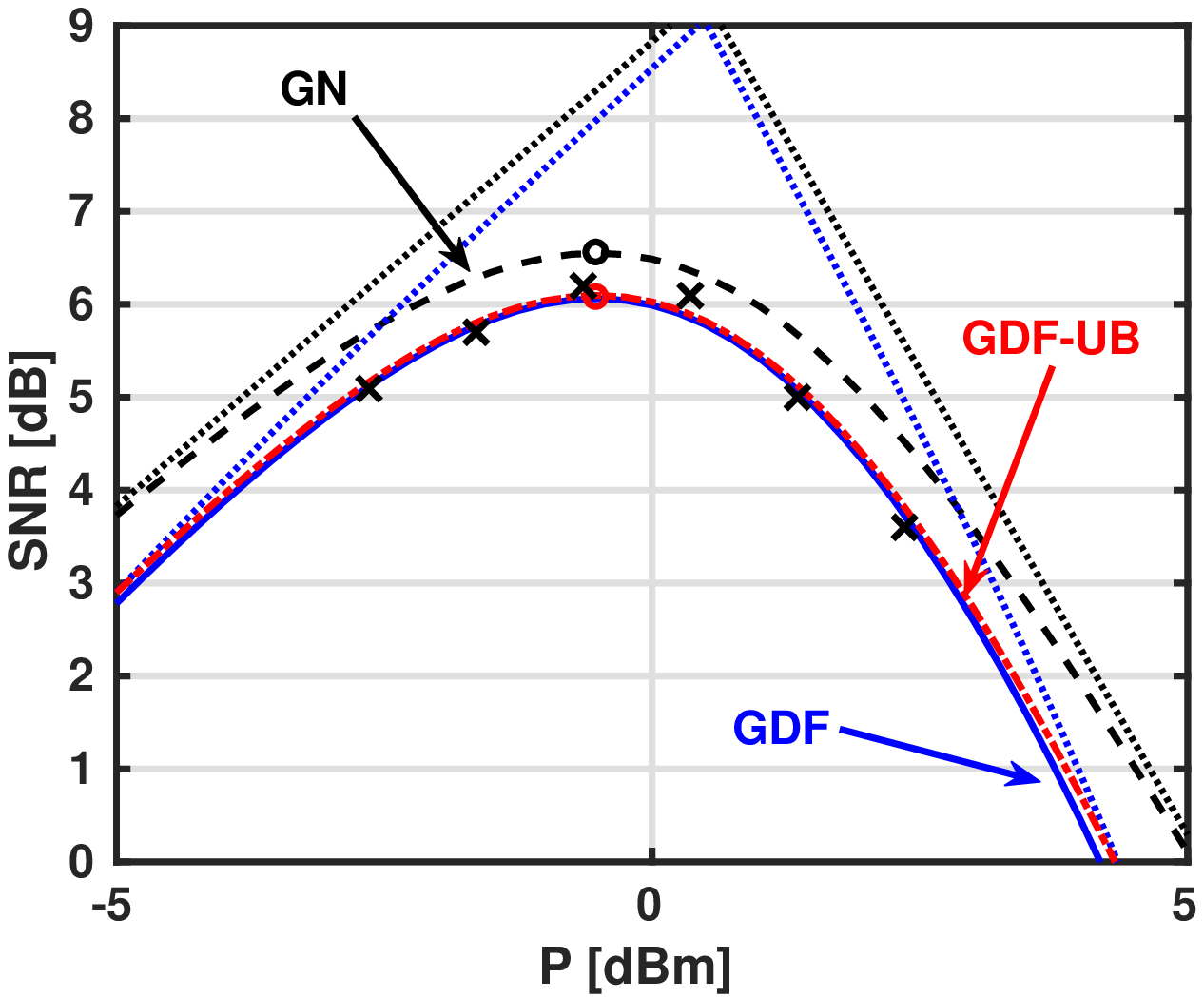}

\label{fig:SNR_vs_P-1}\includegraphics[width=0.85\columnwidth]{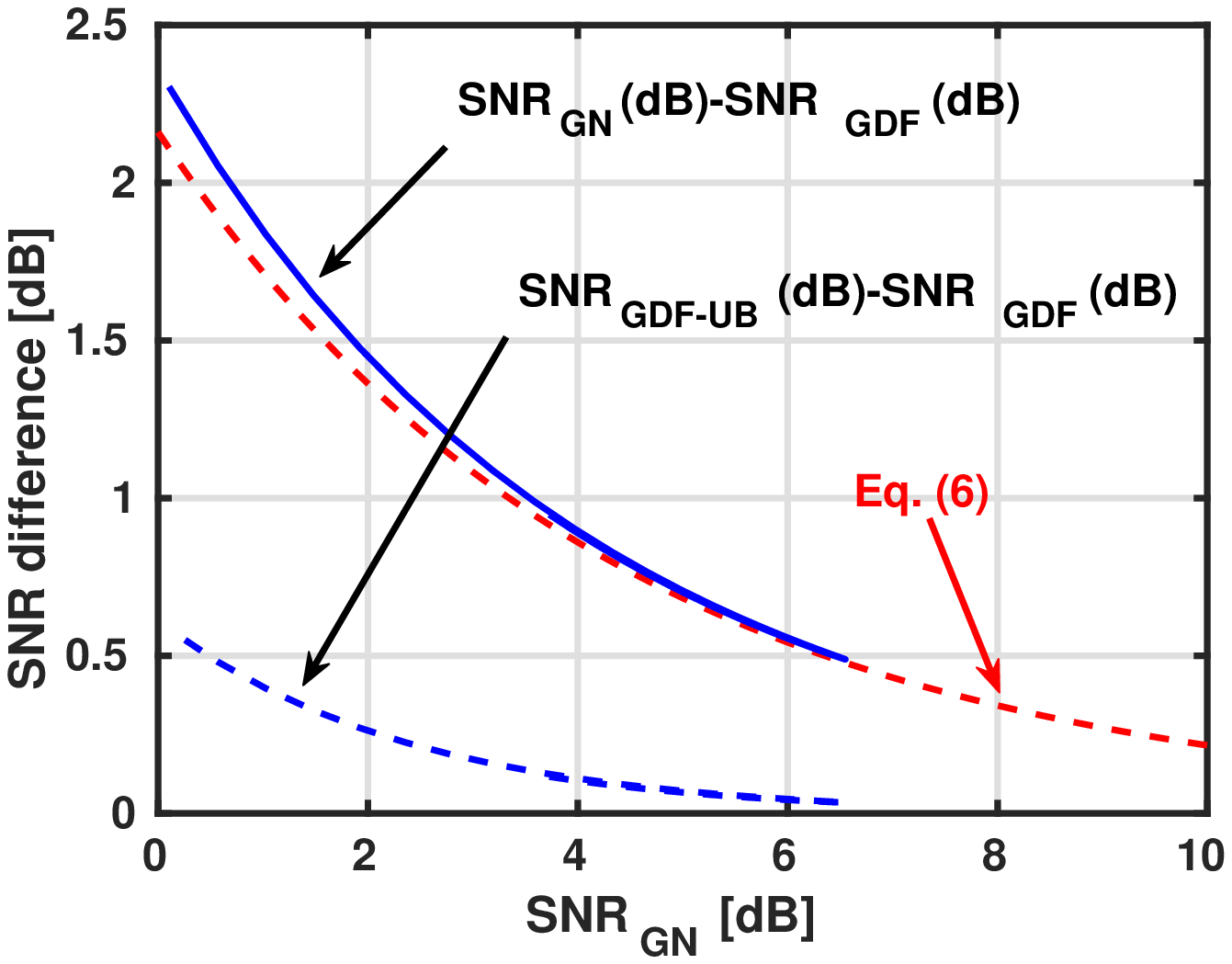}

\caption{\label{fig:SNR_vs_P} \textbf{(top)} nonlinear SNR(dB) versus power per channel
$P$(dBm) for QPSK link in Fig1.a of \cite{ASN_droop_ofc19}, with:
span length $78$ km, fiber loss $0.171$ dB/km, amplifier noise figure
$8$ dB, receiver bandwidth $33$ GHz, $N=228$ spans, $\alpha_{\textrm{NL}}=4.1\times10^{-4}$
mW$^{-2}$. Blue solid: GDF-SNR, eq. (\ref{eq:GDF}). Black dashed:
GN-SNR, eq. (\ref{eq:GN}). Linear and nonlinear asymptotes also shown.
Red dash-dotted: GDF-SNR upper-bound (\ref{eq:GDFapp}). \textbf{(bottom)}
Blue solid line: SNR difference from GN to GDF measured in top figure,
plotted versus GN-SNR. Blue dashed line: GDF-UB to GDF-SNR difference in
top figure, as per eq. (\ref{eq:GDFapp}). Red dashed line:  SNR difference from GN to GDF as per eq. (\ref{eq:SNR_circle-1}).}
\end{figure}

Fig. \ref{fig:SNR_vs_P}(top) shows both the GDF-SNR eq. (\ref{eq:GDF})
(solid line) and the classical GN-SNR (dashed line) plotted versus
power $P$ for the QPSK case reported in \cite[Fig. 1]{ASN_droop_ofc19}.
We note that the top SNR values (circles) for both GDF and GN models
occur at the same optimal power $P_{o}$. However, the tails of the
GDF ``bell curves'' have steeper slopes than their GN counterparts,
as well evidenced by their linear and nonlinear asymptotes. The SNR
gap from GN to GDF in the top figure is plotted as a solid line against
the GN-SNR in Fig. \ref{fig:SNR_vs_P}(bottom). We see that a significant
difference above 0.5dB occurs at all SNRs below 7dB.

Now approximate the total \textcolor{red}{inverse}  droop as

\[
\chi(P)^{\textcolor{red}{-1}}\cong\left(1+\alpha_{\textrm{NL}}P^{2}\right)\left(1+\frac{\beta}{P}\right)\cong1+\frac{\beta}{P}+\alpha_{\textrm{NL}}P^{2}
\]
which is always a reasonable approximation at large enough single-span
linear and nonlinear SNR, namely, $\textrm{SNR}_{1a}\equiv\frac{P}{\beta}$
and $\textrm{SNR}_{1n}\equiv\frac{1}{\alpha_{\textrm{NL}}P^{2}}$. Now let $x\triangleq\frac{\beta}{P}+\alpha_{\textrm{NL}}P^{2}$,
and expand in Taylor to 2nd order $(1+x)^{N}\textcolor{red}{-1}\geq Nx(1+\frac{1}{2}(N-1)x)$.
Then $1/(Nx)\equiv \textrm{SNR}_{\textrm{GN}}$ and we get the upper-bound
\begin{equation}
\textrm{SNR}_{\textrm{GDF}}\leq\frac{\textrm{SNR}_{\textrm{GN}}}{1+\frac{(1-\frac{1}{N})}{2\textrm{SNR}_{\textrm{GN}}}}\triangleq \textrm{SNR}_{\textrm{GDF-UB}}\label{eq:GDFapp}
\end{equation}
valid at any power $P$ and essentially independent of $N$. The upper-bound
$\textrm{SNR}_{\textrm{GDF-UB}}$ in (\ref{eq:GDFapp}) is shown in Fig. \ref{fig:SNR_vs_P}(top)
in red dash-dotted line and is a very good approximation to within 0.5 dB to the true
GDF-SNR, as shown by the blue dashed line SNR difference in Fig. \ref{fig:SNR_vs_P}(bottom).

Also, from (\ref{eq:GDFapp}) we see that
the gap from GN to GDF in dB is approximately $10\log_{10}(1+(1-\frac{1}{N})/(2\textrm{SNR}_{\textrm{GN}}))$,
so that it can be well approximated as 
\begin{equation}
\textrm{SNR}_{\textrm{GN}}(dB)-\textrm{SNR}_{\textrm{GDF}}(dB)\cong\frac{5\log_{10}(e)(1-\frac{1}{N})}{\textrm{SNR}_{\textrm{GN}}}\label{eq:SNR_circle-1}
\end{equation}
as shown in dashed red line in Fig. \ref{fig:SNR_vs_P}(bottom).

\subsection{Optimal power at max SNR}

The optimal power $P_{o}$ at maximum SNR is obtained in the GN model
by setting the derivative of $\textrm{SNR}_{\textrm{GN}}$ w.r.t. $P$ to zero, yielding
the condition $\beta=2\alpha_{\textrm{NL}}P_{o}^{3}$ (i.e., ASE is twice the
NLI at $P_{o}$) and the explicit optimal GN power $P_{o\textrm{GN}}=(\beta/2/\alpha_{\textrm{NL}})^{1/3}$.
Maximum GN-SNR is thus
\begin{equation}
\textrm{SNR}_{o\textrm{GN}}=\frac{1}{N}\frac{1}{3\alpha_{\textrm{NL}}P_{o\textrm{GN}}^{2}}.\label{eq:SNR0_GN}
\end{equation}

Similarly, the GDF-SNR is maximum at the power $P_{o}$ that maximizes
the total droop $\chi(P_{o})$, leading to the condition $\beta=\frac{2}{\chi(P_{o})}\alpha_{\textrm{NL}}P_{o}^{3}$
, ie, ASE is \emph{slightly more than twice} the NLI at $P_{o}$.
This leads to $P_{o}=P_{o\textrm{GN}}\chi^{1/3}\lesssim P_{o\textrm{GN}}$, thus \emph{the
optimal $P_{o}$ is in practice the same as in the GN case}. Using
the bound (\ref{eq:SNR_circle-1}), starting from the top GN-SNR we computed the red circle in
Fig. \ref{fig:SNR_vs_P}, which falls right on top of the maximum
of the GDF-SNR curve. Fig. \ref{fig:SNR_vs_P}(bottom, red-dashed line)
shows that predicting the GDF-SNR using the upper-bound (\ref{eq:GDFapp})
based on the GN-SNR is accurate down to SNR values as low as 0 dB.

\subsection{Spectral efficiency}

A lower bound on the capacity per mode of the nonlinear optical channel for
dual-polarization  transmissions is obtained from the equivalent
additive white Gaussian noise channel (AWGN) Shannon capacity, i.e.,
by considering the NLI as an additive white Gaussian process independent
of the signal. Hence a lower bound on spectral efficiency per mode is \cite{poggiotutto}:
$\textrm{SE}=2\log_{2}(1+\textrm{SNR})$. Its top value $\textrm{SE}_{o}$ is achieved at $P_{0}$.
\begin{figure}
\centering\includegraphics[width=0.85\columnwidth]{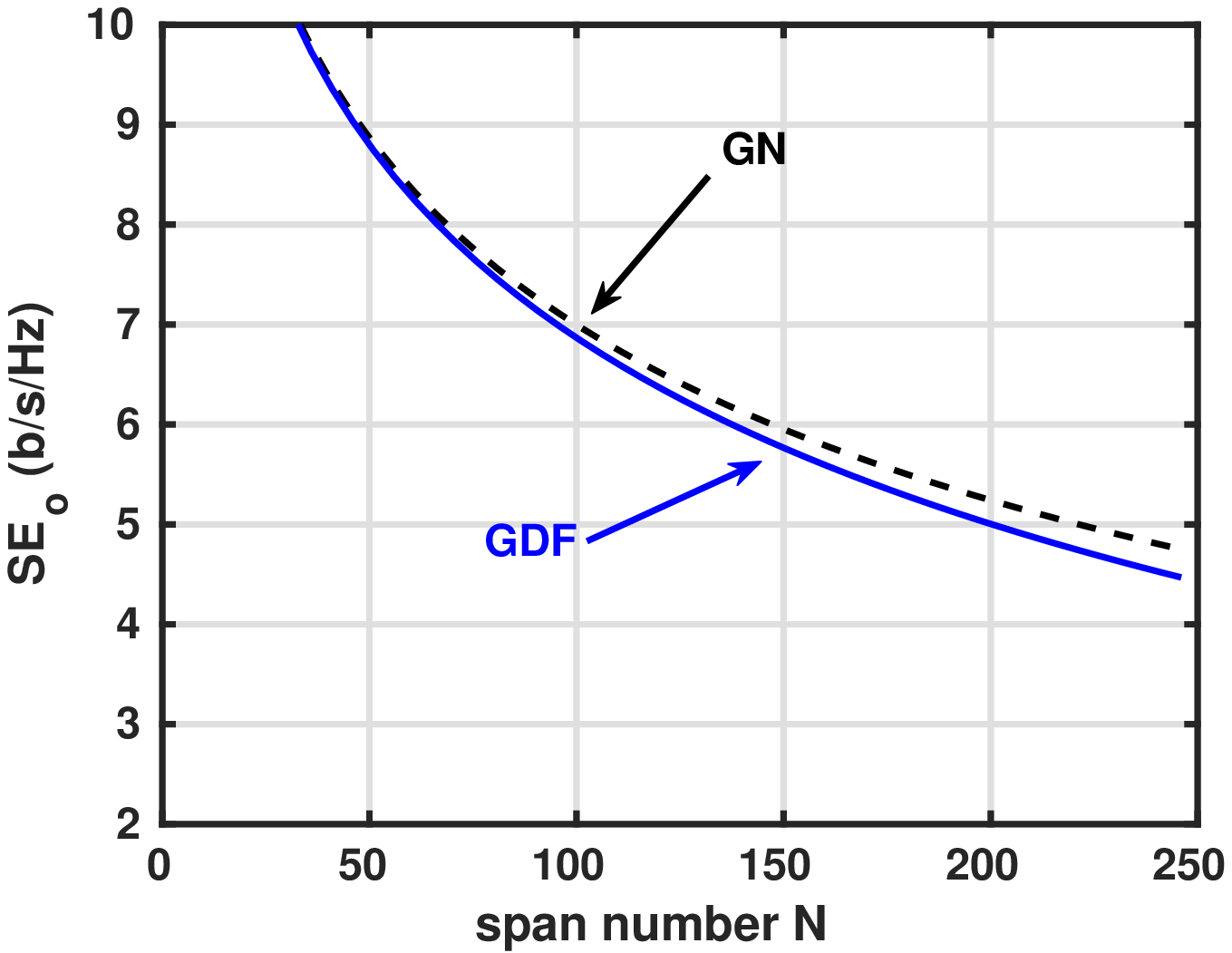}

\includegraphics[width=0.85\columnwidth]{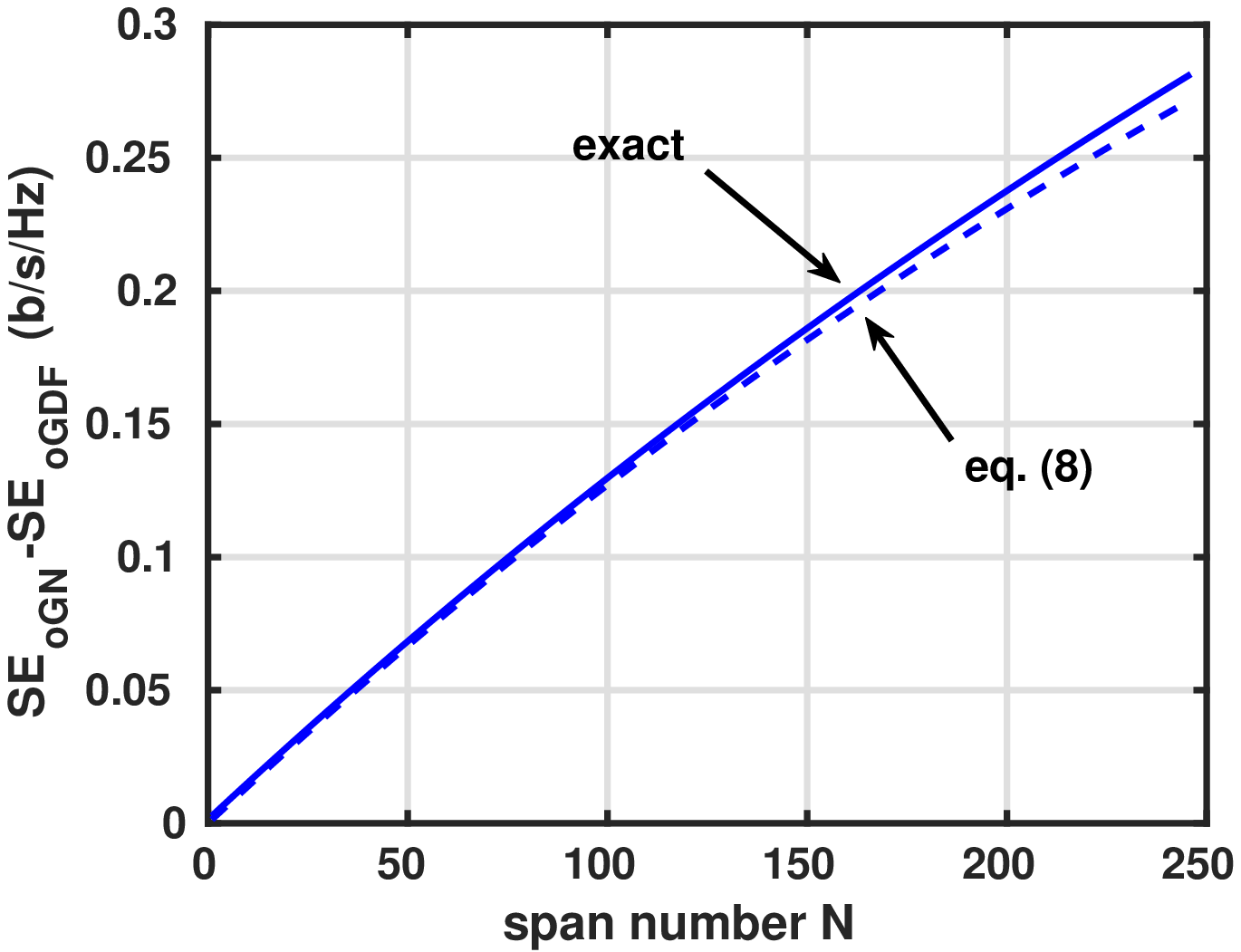}

\caption{\label{fig:SEvsN} \textbf{(top)} Maximum Spectral efficiency per mode versus span number $N$.
Dashed: GN model. Solid: GDF. Same parameters as in Fig. \ref{fig:SNR_vs_P}(\textcolor{red}{top}).
\textbf{(bottom)} top SE difference between GN and GDF. Solid: exact; dashed: eq
(\ref{eq:Diff_SE}).}
\end{figure}

Fig. \ref{fig:SEvsN}(top) reports $\textrm{SE}_o$ versus number of spans $N$
for both GN and GDF using the same data as in Fig. \ref{fig:SNR_vs_P}.
We note that a significant departure from the GN formula occurs only
at spectral efficiencies smaller than 5 b/s/Hz. We find that the $\textrm{SE}_o$
gap from GN to GDF is well approximated by the formula 
\begin{equation}
\textrm{SE}_{o\textrm{GN}}-\textrm{SE}_{o\textrm{GDF}}\cong\frac{2}{\ln(2)}\frac{\textrm{SNR}_{o\textrm{GN}}}{1+2\textrm{SNR}_{o\textrm{GN}}+2\textrm{SNR}_{o\textrm{GN}}^{2}}\label{eq:Diff_SE}
\end{equation}
which is plotted in Fig. \ref{fig:SEvsN}(bottom) together with the
exact gap. It is seen that the GN model over-estimates $\textrm{SE}_o$ by less than 0.3
b/s/Hz at the longest distance.

\section{Conclusions}

We have presented an analytical model that fully justifies the generalized
droop formula for SNR. We quantified its difference from the GN model
SNR, which, for the analyzed link, becomes larger than 0.5 dB  only at SNR values below about 7dB, typical of submarine links with hundreds of amplifiers. We note
that the NLI droop derivation extends verbatim to any power-redistributing
effect in the fiber, such as for instance the GAWBS.


\section{References}

\begin{thebibliography}{14}

\bibitem{ASN_droop_ofc19} Antona, J.,
Meseguer, A. C. , and Letellier, V.:  \textquoteleft Transmission Systems
with Constant Output Power Amplifiers at Low SNR Values: a Generalized
Droop Model\textquoteright, in Optical Fiber Communication Conference
(OFC) 2019, paper M1J.6. 

\bibitem{asn_suboptic} Antona, J.C., Dupont,  S., Poullias, G.,   et al.: \textquoteleft Performance of open cable: from modeling to wide scale experimental assessment\textquoteright, in Proc. SubOptic, April 8-12, 2019, New Orleans (LA).

\bibitem{bergano} Bergano, N. : \textquoteleft Undersea Communications
Systems\textquoteright , in \emph{Optical Fiber Telecommunications
IV B Systems and Impairments}, I. P. Kaminow and T. Li Eds. Academic
Press, San Diego (CA), 2002.

\bibitem{subcom}\small Sinkin, O. V., Turukhin, A. V., Sun, Y.,
et al.: \textquoteleft SDM for Power-Efficient Undersea Transmission\textquoteright ,
J. Lightw. Technol., 2018, {\bf 36}, (2), pp. 361-371.

\bibitem{poggioPTL11}\small Poggiolini, P., Carena, A., Curri, V.,
et al.: \textquoteleft Analytical Modeling of Non-Linear Propagation
in Uncompensated Optical Transmission Links\textquoteright , Photon.
Technol. Lett., 2011, {\bf 23}, (11), pp. 742-744.

\bibitem{grellierOptExp}\small E. Grellier and A. Bononi: \textquoteleft Quality
parameter for coherent transmissions with Gaussian-distributed nonlinear
noise\textquoteright , Opt. Express, 2011, {\bf 19}, (13), pp.
12781--12788. 

\bibitem{poggio_depletion-2}P. Poggiolini, A. Carena, Y. Jiang et
al.: \textquoteleft Impact of Low-OSNR Operation on the Performance
of Advanced Coherent Optical Transmission Systems\textquoteright ,
in European Conference on Optical Communications
(ECOC) 2014, paper Mo.4.3.2.

\bibitem{poggiotutto}Poggiolini, P. , Bosco, G., Carena, A. et al.: \textquoteleft The GN-Model of Fiber Non-Linear Propagation and its Applications\textquoteright ,
J. Lightw. Technol., 2014, {\bf 32}, (4), pp. 694-721.

\bibitem{GAWBS}M. A. Bolshtyansky, J.-X. Cai, C. R. Davidson et al.:
\textquoteleft Impact of Spontaneous Guided Acoustic-Wave Brillouin
Scattering on Long-haul Transmission\textquoteright , in Optical Fiber
Communication Conference (OFC) 2018, paper M4B.3

\end{thebibliography}
\end{document}